\documentclass[a4paper,11pt]{article}
\usepackage{pos}
\usepackage{bibspacing}

\newcommand{\tcite}[1]{~\cite{#1}}
\newcommand{\tref}[1]{~\ref{#1}}
\newcommand{\eref}[1]{~\eqref{#1}}

\title{Toward twist-2 $T$-odd transverse-momentum-dependent gluon distributions: the $f$-type Sivers function
}
\ShortTitle{Toward twist-2 $T$-odd TMD gluon distributions: the Sivers function}

\author[a,b]{Alessandro Bacchetta}
\author*[c,d,e]{Francesco Giovanni Celiberto}
\author[b]{Marco Radici}

\affiliation[a]{Dipartimento di Fisica, Universit\`a di Pavia, via Bassi 6, I-27100 Pavia}
\affiliation[b]{INFN Sezione di Pavia, via Bassi 6, I-27100 Pavia, Italy}
\affiliation[c]{European Centre for Theoretical Studies in Nuclear Physics and Related Areas (ECT*), I-38123 Villazzano, Trento, Italy}
\affiliation[d]{Fondazione Bruno Kessler (FBK), I-38123 Povo, Trento, Italy}
\affiliation[e]{INFN-TIFPA Trento Institute of Fundamental Physics and Applications, I-38123 Povo, Trento, Italy}

\emailAdd{alessandro.bacchetta@unipv.it}
\emailAdd{fceliberto@ectstar.eu}
\emailAdd{marco.radici@pv.infn.it}

\abstract{
We report progresses on leading-twist transverse-momentum-dependent (TMD) gluon distributions, calculated in a spectator-model framework for the parent nucleon. We encode in our formalism a fit-based parameterization for the spectator-mass density, suited to describe the gluon content of the proton in a wide range of $x$. We present preliminary results on the $f$-type Sivers function, a key ingredient to access relevant spin-asymmetries emerging when the nucleon is transversely polarized. These studies are helpful to shed light on the gluon dynamics inside nucleons and nuclei, which is one of the primary goals of new-generation colliding facilities, as the Electron-Ion Collider, the High-Luminosity LHC, NICA-SPD, and the Forward Physics Facility.
}

\FullConference{%
  *** The European Physical Society Conference on High Energy Physics (EPS-HEP2021), ***\\
  *** 26-30 July 2021 ***\\
  *** Online conference, jointly organized by Universität Hamburg and the research center DESY ***
}

\begin{document}
\maketitle

\section{Introduction}

The search for clues of New Physics is in the viewfinder of current and forthcoming analyses at the Large Hadron Collider (LHC) and at new-generation colliders. At the same time, a window of opportunities is opened to enhance our knowledge of strong interactions and, notably, of the dynamics of hadrons' constituents in terms of parton densities and fragmentation functions. 
A key role in answering fundamental questions of Quantum Chromodynamics (QCD), such as the origin of proton mass and spin, is played by our ability of reconstructing the three-dimensional motion of partons inside parent nucleons. This \emph{tomographic} description is naturally afforded by the so-called transverse-momentum dependent (TMD) factorization formalism.
Though significant results in recent years were collected for quark TMD densities, the deep knowledge on their formal properties being corroborated by encouraging phenomenological results, the gluon TMD sector still represents an almost uncharted field. Polarized gluon TMD distributions were classified for the first time in Ref.\tcite{Mulders:2000sh} and then in Refs.\tcite{Meissner:2007rx,Lorce:2013pza,Boer:2016xqr}, while pioneering phenomenological analyses were proposed in Refs.\tcite{Lu:2016vqu,Lansberg:2017dzg,Gutierrez-Reyes:2019rug,Scarpa:2019fol,Adolph:2017pgv,DAlesio:2017rzj,DAlesio:2018rnv,DAlesio:2019qpk}.
A striking evidence that, at variance with their collinear counterparts, TMD densities are process-dependent emerged from theoretical studies on \emph{spin asymmetries}\tcite{Brodsky:2002cx,Collins:2002kn,Ji:2002aa}.
From the formal point of view, the TMD process dependence is obtained by following the color flow over the integration path of \emph{gauge links}.
It turns out that different classes of processes probe distinct gluon
gauge-link structures, thus resulting in a more diversified kind of \emph{modified universality} with respect to what happens for the quark case.
This leads to a distinction between two major gluon gauge links: the $f\text{-type}$ and the $d\text{-type}$ one, respectively known in the context of small-$x$ analyses as Weisz\"acker--Williams and dipole structures~\cite{Kharzeev:2003wz,Dominguez:2010xd,Dominguez:2011wm}.
The antisymmetric $f_{abc}$ QCD color structure directly enters the analytic expression for the $f$-type $T$-odd gluon-TMD correlator, while the symmetric $d_{abc}$ structure appears in the $d$-type $T$-odd one. The $[+,+]$ and $[-,-]$ gauge links are the building blocks of the $f$-type TMDs, while the $[+,-]$ and $[-,+]$ ones distinguish the $d$-type ones\footnote{The $+$ and $-$ links indicate the direction of future- and past-pointing Wilson lines corresponding to final- and initial-state interactions, respectively.}.
More convoluted, box-loop gauge links appear in processes where multiple color states connects both the initial and final state~\cite{Bomhof:2006dp}, this bringing to factorization-violation effects~\cite{Rogers:2013zha}.
In the high-energy factorization regime (large transverse momentum and small-$x$), the unpolarized and linearly polarized gluon TMDs, $f_1^g$ and $h_1^{\perp g}$ collapse~\cite{Dominguez:2011wm} to the unintegrated gluon distribution (UGD), genuinelly defined inside the BFKL formalism~\cite{Fadin:1975cb,Balitsky:1978ic} (see Refs.~\cite{Hentschinski:2012kr,Besse:2013muy,Bolognino:2018rhb,Bolognino:2018mlw,Bolognino:2019bko,Bolognino:2019pba,Celiberto:2019slj,Brzeminski:2016lwh,Garcia:2019tne,Celiberto:2018muu,Celiberto:2020wpk,Celiberto:2015yba,Bolognino:2019yls,Bolognino:2021niq,Celiberto:2020tmb,Celiberto:2020rxb,Bolognino:2021mrc,Celiberto:2021dzy,Celiberto:2021fdp,Celiberto:2022dyf} for quite recent applications).

With the aim of providing the scientific community with a flexible model, suited to phe\-no\-me\-no\-lo\-gy, a general framework was recently set up\tcite{Bacchetta:2020vty} (see also Refs.\tcite{Bacchetta:2021oht,Celiberto:2021zww,Celiberto:2022fam,Bacchetta:2021lvw,Bacchetta:2021twk,Bacchetta:2022esb}) for all the leading-twist $T$-even gluon TMDs, calculated in a spectator model for the parent proton and embodying effective small-$x$ BFKL-resummation effects.
In this work we present recent developments toward $T$-odd functions, and in particular to the calculation of the $f$-type Sivers TMD, which is directly connected to the density of unpolarized gluons inside transversely polarized nucleons.

\section{The $f$-type Sivers function in a spectator model}

\begin{figure}[t]
 \centering

 \includegraphics[width=0.38\textwidth]{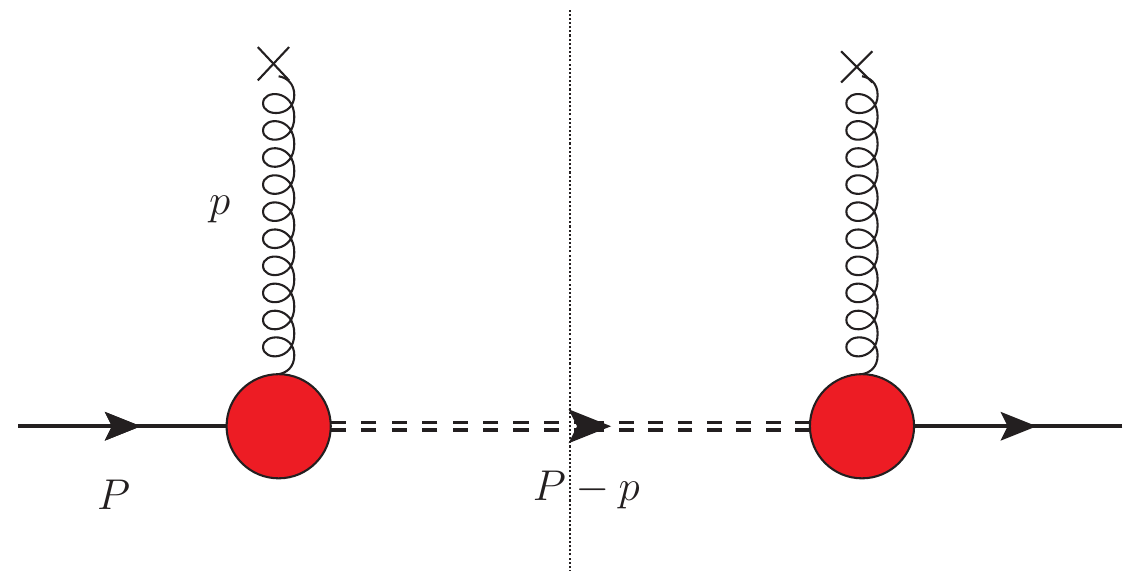} \hspace{1.00cm}
 \includegraphics[width=0.43\textwidth]{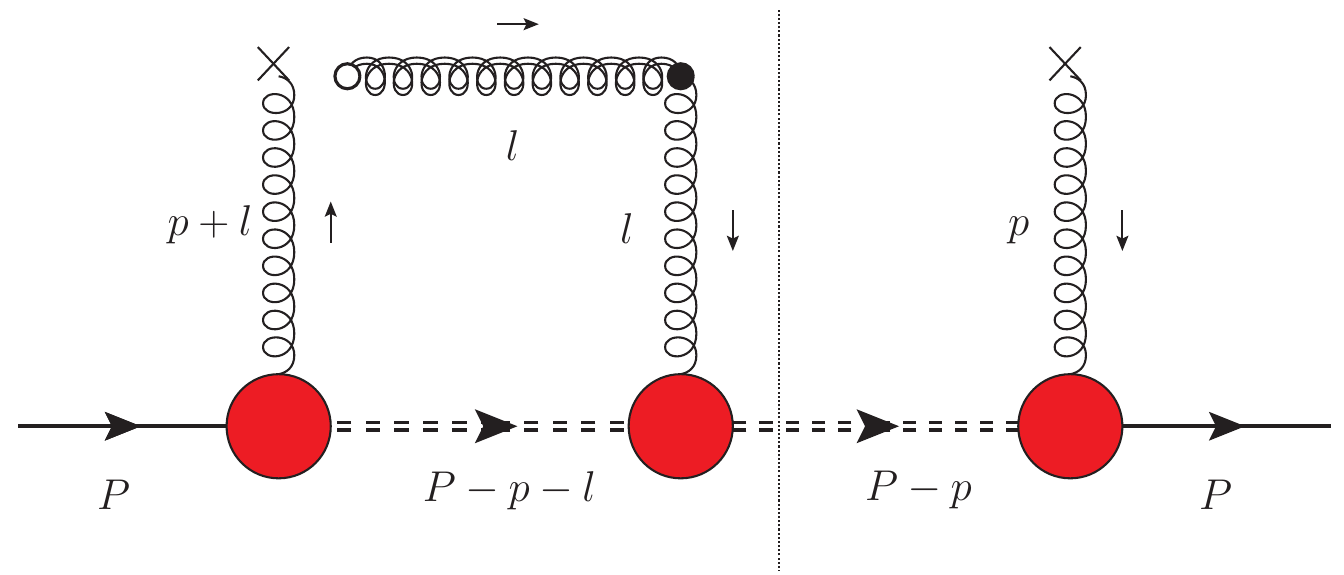}

\caption{
Tree-level cut diagram for the calculation of $T$-even twist-2 gluon TMDs (left), and lowest-order diagram for $T$-odd $[+,+]$ ones (right, Hermitian-conjugate diagram not shown).
Red blobs represent nucleon-gluon-spectator effective vertices, dashed double line stands for spin-$\textstyle{\frac{1}{2}}$ spectators, and the curly double line depicts a gluon eikonal propagator.}

\label{fig:correlator}
\end{figure}

The main assumption of the spectator-model framework is that the nucleon with mass $M$ and four-momentum $P$ can emit a parton having four-momentum $p$, longitudinal fraction $x$ and transverse momentum $\boldsymbol{p}_T$, while remainders are effectively treated as a particle with mass $M_X$, called spectator.
Our choice for nucleon-gluon-spectator vertex reads
\begin{equation}
 \label{eq:vertex}
 {\cal Y}^\mu_{ab} = \delta_{ab} \left( g_1(p^2)\gamma^\mu + g_2(p^2) \frac{i}{2M} \sigma^{\mu\nu}p_\nu \right) \, ,
\end{equation}
the $g_{1,2}$ functions being two dipolar functions of $\boldsymbol{p}_T^2$. The use of dipolar form factors allows us to remove gluon-propagator divergences, dampen large-$\boldsymbol{p}_T$ effects that cannot be caught by a pure TMD description, and eliminate logarithmic singularities arising from $\boldsymbol{p}_T$-integrated distributions.
A comprehensive study on spectator-model quark TMDs in the proton was done in Refs.\tcite{Bacchetta:2008af,Bacchetta:2010si}, by taking into account different spin states for di-quark spectators and several nucleon-parton-spectator form factors.
In Ref.\tcite{Bacchetta:2020vty} all twist-2 $T$-even gluon TMDs in the proton were calculated via an improved version of the tree-level spectator model (see left panel of Fig.\ref{fig:correlator}), where the spectator was treated as a colored spin-1/2 particle and its mass $M_X$ was allowed to be in a range of values weighted by the 7-parameter spectral function given below
\begin{equation}
\label{eq:rhoX}
 \rho_{\rm } (M_X) = \mu^{2a} \left( \frac{A}{B + \mu^{2b}} + \frac{C}{\pi \sigma} e^{-\frac{(M_X - D)^2}{\sigma^2}} \right) \,.
\end{equation}
Model parameters were fixed by performing a simultaneous fit of our unpolarized and helicity TMDs, $f_1^g$ and $g_1^g$, to the corresponding collinear PDFs from {\tt NNPDF}\tcite{Ball:2017otu,Nocera:2014gqa} at the initial scale $Q_0 = 1.64$ GeV. The statistical uncertainty of the fit is obtained through the well known bootstrap method.
We remark that our tree-level approximation does not account for the gauge link. Therefore, our model $T$-even functions are process-independent.

\begin{figure}[h]
\centering
\includegraphics[width=0.42\textwidth]{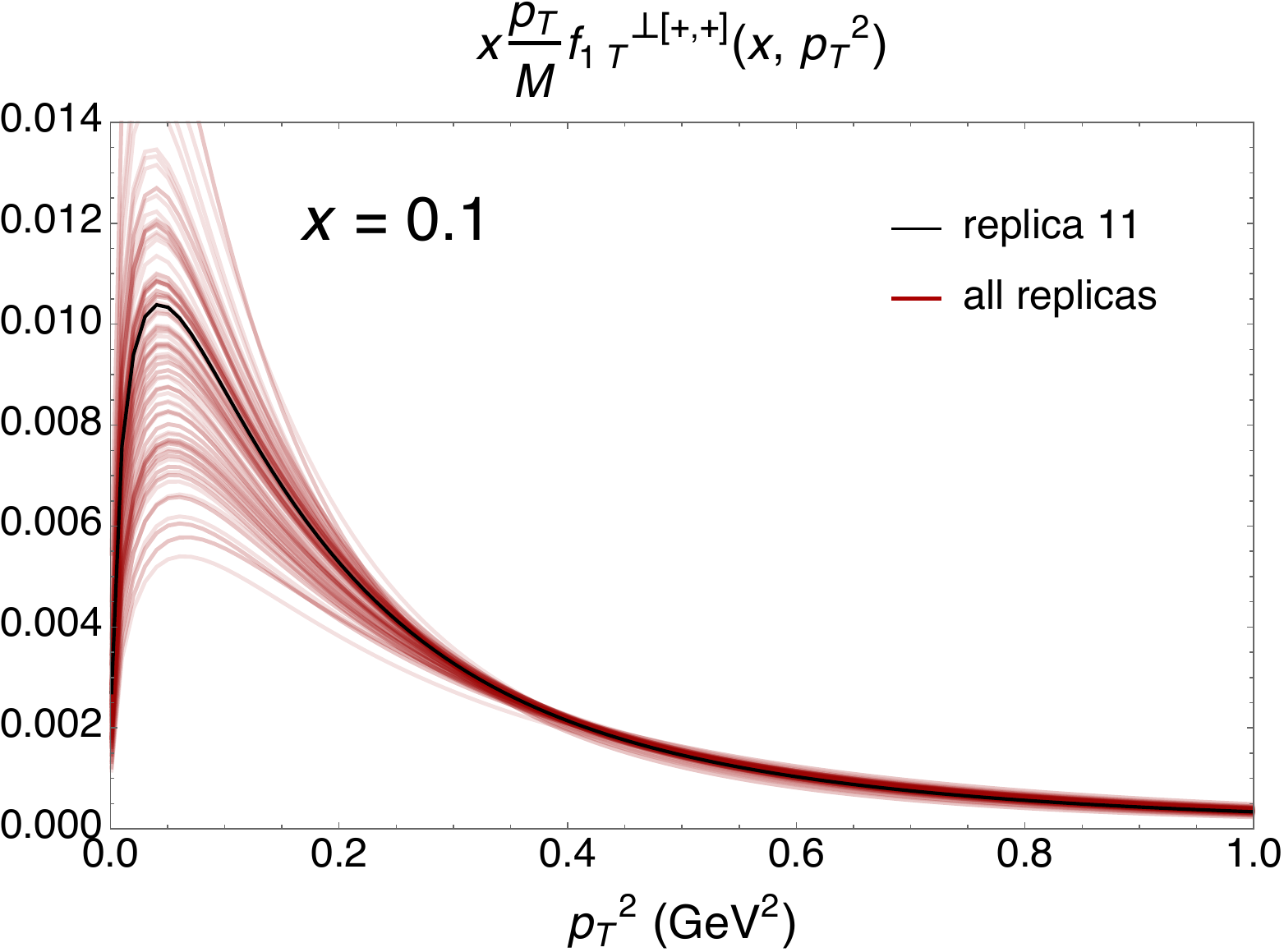} \hspace{0.75cm}
\includegraphics[width=0.42\textwidth]{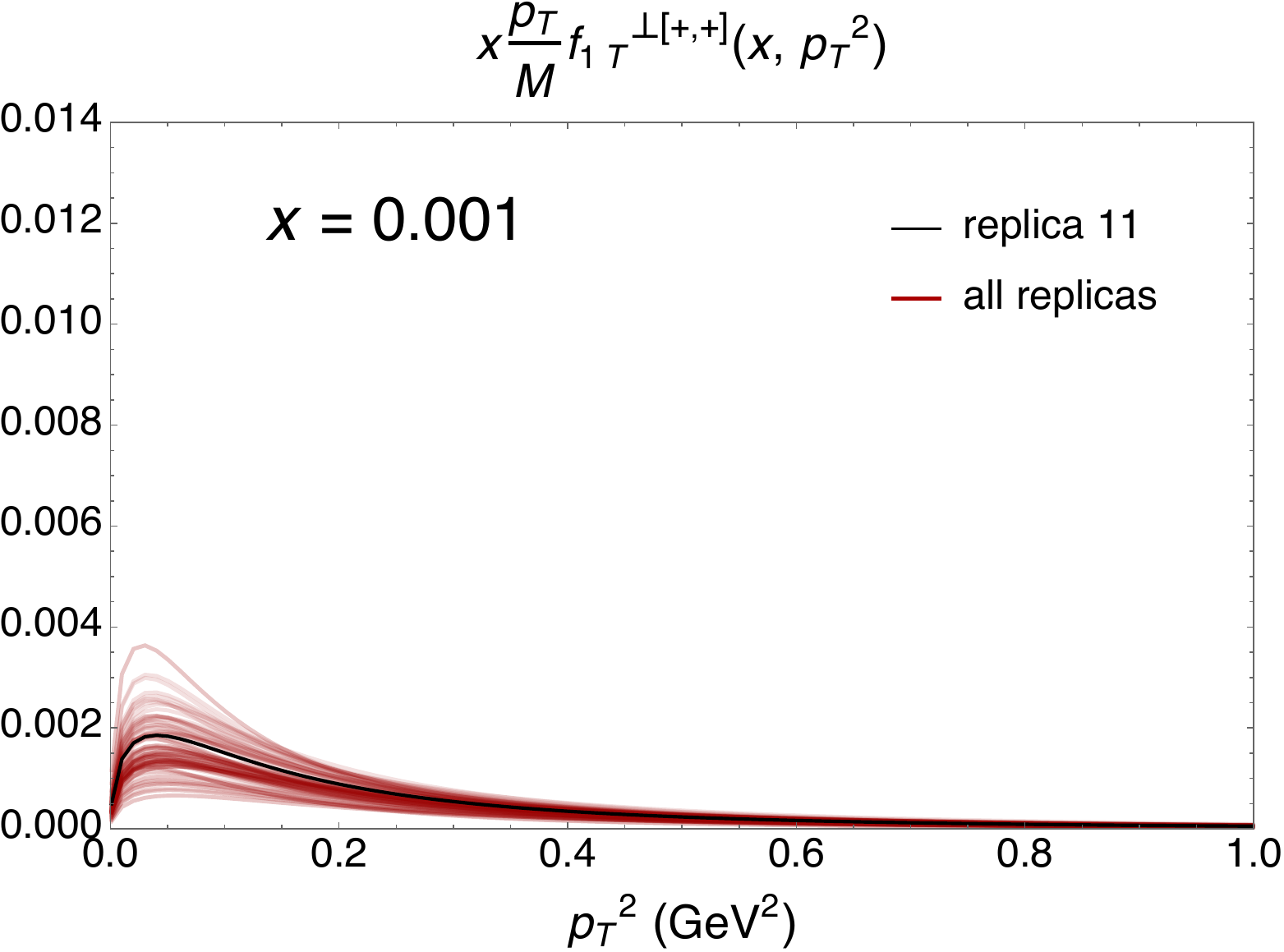}
\caption{Transverse-momentum dependence of the $[+,+]$ Sivers function calculated in the spectator model, for $x=10^{-1}$ (left) and $x=10^{-3}$ (right), and at the initial scale $Q_0 = 1.64$ GeV. Black curve portrays the most representative replica \#11.}
\label{fig:Sivers_f}
\end{figure}

A naive calculation of any $T$-odd TMD via the tree-level gluon correlator would lead to a vanishing result. It is due to fact that at the tree-level one has just a real scattering amplitude, while, in order to have a nonzero $T$-odd function, a phase factor coming from the interference between two scattering amplitudes with different imaginary parts is needed. One can generate this factor by considering the interference with one-gluon exchange in the \emph{eikonal} approximation, as in the right panel of Fig.\eref{fig:correlator}. Since this corresponds to introducing the leading-twist
one-gluon-exchange approximation of the gauge link operator, the calculated $T$-odd TMDs turn out to be process-dependent. For a given choice of the gauge link, say the $f\text{-type}$ one, the two corresponding Sivers functions are obtained via the following projection
\begin{equation}
 \frac{\epsilon_T^{pS}}{M} f_{1T}^{\perp g \, [+,+]}(x, \boldsymbol{p}_T^2) \equiv - \frac{\epsilon_T^{pS}}{M} f_{1T}^{\perp g \, [-,-]}(x, \boldsymbol{p}_T^2) = \frac{1}{2} g_T^{ij} \left[ \Phi_{i j}^f (x, \boldsymbol{p}_T, S_T) -  \Phi_{i j}^f (x, \boldsymbol{p}_T, -S_T) \right] \, ,
\label{eq:Sivers_f}
\end{equation}
with $\Phi_{i j}^f$ the $f$-type gluon correlator, $S_T$ the transverse polarization of the proton, $g_T^{ij}$ the transverse metric tensor, and $\epsilon_T^{pS} \equiv \epsilon^{-+ij} p_i S_j$.
In our preliminary analysis we rely on a simplified version of the nucleon-gluon-spectator vertex, namely we set the $g_2$ form factor in Eq.\eref{eq:vertex} at zero.
For the sake of consistency, we use new model parameters fitted to {\tt NNPDF} results with this approximated expression for the vertex.
We present in Fig.\tref{fig:Sivers_f} the $\boldsymbol{p}_T^2$-shape of the $[+,+]$ Sivers function at $x = 10^{-1}$ and at $x = 10^{-3}$, and for the initial scale $Q_0 = 1.64$ GeV.
We clearly observe a non-Gaussian pattern in $\boldsymbol{p}_T^2$, a large flattening tail in the $\boldsymbol{p}_T^2 \to 1$ GeV limit, and a small nonzero value when $\boldsymbol{p}_T^2 \to 0$.
At variance with the unpolarized and the Boer--Mulders gluon TMDs (see Fig.~(4) of Ref.\tcite{Bacchetta:2020vty}), the $f$-type Sivers function exhibits a decreasing behavior with $x$. This suggests that transverse single-spin asymmetries could be more evident in the moderate-$x$ regime.
We stress, however, that our results could change even radically when the full-vertex calculation will be available.

\vspace{-0.60cm}
\section{Summary and Outlook}

We have included in our spectator-model framework the calculation of the $f$-type gluon Sivers TMD function with a simplified expression for the nucleon-gluon-spectator vertex. The extension to the full-vertex case is underway, as well as the calculation of the remnant three $T$-odd densities at twist-2 (linearity, pretzelosity and pseudo worm-gear) and of the corresponding $d$-type ones.
All these studies represent a useful guidance toward the exploration of observables sensitive to gluon-TMD dynamics at new-generation colliding facilities, such as the \emph{Electron-Ion Collider}~(EIC)~\cite{AbdulKhalek:2021gbh}, the \emph{High-Luminosity Large Hadron Collider} (HL-LHC)~\cite{Chapon:2020heu}, NICA-SPD~\cite{Arbuzov:2020cqg}, and the \emph{Forward Physics Facility} (FPF)~\cite{Anchordoqui:2021ghd}.

\vspace{-0.25cm}
\bibliographystyle{bibstyle}
\bibliography{references}

\end{document}